\documentclass[twocolumn]{aastex62}
\usepackage{color}
\usepackage[space]{grffile}
\usepackage{latexsym}
\usepackage{textcomp}
\usepackage{longtable}
\usepackage{multirow,booktabs}
\usepackage{amsfonts,amsmath,amssymb}
\usepackage{url}
\usepackage{hyperref}
\usepackage[english]{babel}
\usepackage{graphicx}
\usepackage[space]{grffile}
\usepackage{url}
\usepackage[utf8]{inputenc}
\usepackage{hyperref}
\usepackage{longtable}
\usepackage{natbib}

\newcommand\avg[1]{\left\langle{#1}\right\rangle}

\begin{document}

\shorttitle{Ultra-faint Satellites} 
\shortauthors{F. Munshi et al.}
\title{Dancing in the Dark: Uncertainty in ultra-faint dwarf galaxy predictions from cosmological simulations}
\author{Ferah Munshi}
\affiliation{Department of Physics \&
Astronomy, University of Oklahoma\\ 440 W. Brooks St., Norman, OK 73019}
\affiliation{Department of Physics \&
Astronomy, Vanderbilt University\\ 6301 Stevenson Center, Vanderbilt University, Nashville, TN 37235 }
\affiliation{Department of Physics \&
Astronomy, Rutgers, The State University of New Jersey\\ 136 Frelinghuysen Rd, Piscataway, NJ 08854}

\author{Alyson M. Brooks}
\affiliation{Department of Physics \&
Astronomy, Rutgers, The State University of New Jersey\\ 136 Frelinghuysen Rd, Piscataway, NJ 08854}

\author{Charlotte Christensen}
\affiliation{Department of Physics \&
Astronomy, Grinnell University\\ Grinnell, IA 50112}

\author{Elaad Applebaum}
\affiliation{Department of Physics \&
Astronomy, Rutgers, The State University of New Jersey\\ 136 Frelinghuysen Rd, Piscataway, NJ 08854}

\author{Kelly Holley-Bockelmann}
\affiliation{Department of Physics \&
Astronomy, Vanderbilt University\\ 6301 Stevenson Center, Vanderbilt University, Nashville, TN 37235 }

\author{Thomas R. Quinn}
\affiliation{Department of 
Astronomy, University of Washington\\  3910 15th Ave., Seattle, WA 98195}
\author{James Wadsley}
\affiliation{Department of Physics \&
Astronomy, McMaster University\\ Hamilton, Ontario L8S 4K1 }




\begin{abstract}
The existence of ultra-faint dwarf (UFD) galaxies highlights the need to push our theoretical understanding of galaxies to extremely low mass. We examine the formation of UFDs by twice running a fully cosmological simulations of dwarf galaxies, but varying star formation. One run uses a temperature-density threshold for star formation, while the other uses an H$_{2}$-based sub-grid star formation model. The total number of dwarf galaxies that forms is different by a factor of 2 between the two runs, but most of these are satellites, leading to a factor of 5 difference in the number of luminous UFD companions around more massive, isolated dwarfs.  The first run yields a 47\% chance of finding a satellite around a M$_{halo}$ $\sim 10^{10}$ M$_{\odot}$ host, while the H$_2$ run predicts only a 16\% chance.  Metallicity is the primary physical parameter that creates this difference.  As metallicity decreases, the formation of H$_2$ is slowed and relegated to higher-density material.  Thus, our H$_2$ run is unable to form many (and often, any) stars before reionization removes gas.  These results emphasize that predictions for UFD properties made using hydrodynamic simulations, in particular regarding the frequency of satellites around dwarf galaxies, the slope of the stellar mass function at low masses, as well as the properties of ultra-faint galaxies occupying the smallest halos, are extremely sensitive to the subgrid physics of star formation contained within the simulation. However, upcoming discoveries of ultra-faint dwarfs will provide invaluable constraining power on the physics of the first star formation.

\end{abstract}


\section{Introduction}
 How do galaxies populate low mass dark matter halos?  Is there a lower limit to the halo mass that can form a galaxy?  The lowest mass galaxies that we have observed are the ultra-faint dwarf (UFD) galaxies, which are typically defined to have M$_{star} \lesssim$ 10$^5$ M$_{\odot}$.  The lowest mass galaxy yet observed contains only a few hundred M$_{\odot}$ in stars \citep{Homma2018}.  To understand these extremely low mass systems, a few authors have simulated the formation of isolated UFD galaxies at high resolution \citep{Read2016,Jeon2017,Corlies2018}.  Some cosmological simulations of classical dwarf galaxies have been able to resolve the formation of UFD satellite companions \citet{Wheeler2015,Wheeler2018,Munshi2017}. In general, fully cosmological simulations of more massive galaxies like the Milky Way have not had sufficient resolution to resolve the formation of UFD companions, preventing direct predictions for UFD abundances and distributions that can be tested with the Large Synoptic Survey Telescope \citep[LSST;][]{Walsh2009,tollerud08}. However, the next generation of cosmological simulations will achieve stellar mass resolutions of $\sim$1000 M$_{\odot}$, allowing simulators to start pushing down into the UFD range.  In this paper, we explore whether the prescriptions commonly adopted by simulators to create realistic Milky Way-mass and classical dwarf galaxies can be extrapolated down to UFD scales and the impact of prescription choice on observational predictions made using cosmological simulations. 

 Cosmological simulations of galaxies run to $z=0$ have  recently been quite successful in matching a long list of observed scaling relations \citep{G10, Brook2012, G12,Aumer2013, Munshi13, Vogelsberger2013b,Shen2014,BZ2014, FIRE, NIHAO, EAGLE, Sawala2016b, Christensen2016, Garrison-Kimmel2018, Santos-Santos2018, Nelson2018}. This success is despite the fact that different simulators often adopt different physical prescriptions, particularly the prescriptions for star formation and energetic feedback from stars, supernovae, and AGN. It is generally agreed that different simulators can broadly reproduce the same observed trends despite different prescriptions because galaxies ``self-regulate,'' i.e., a change in the star formation prescription is counter-balanced by subsequent feedback \citep{saitoh08, Hopkins2011,Hopkins2013c, Christensen2014b, Benincasa2016,FIRE2,Pallottini2017}.  Previous studies found that self-regulation can occur as long as the resolution is high enough to capture the average densities in giant molecular clouds (GMCs), and therefore that the simulation is high enough resolution to have star formation limited to the scales of GMCs \citep{Agertz2016,Semenov2016,Buck2019}.  

\citet{Semenov2018} show that self-regulation is limited to the regime of strong feedback, which regulates the gas supply available to turn into stars.  Star formation efficiency drops as halo mass decreases, so it is not clear that UFDs lie in the regime of strong feedback, or have the ability to self-regulate.  In fact, some simulators have shown that low mass halos can completely shut off their own star formation via feedback \citep[e.g.,][]{Fitts2017,Wright2018}.  UFDs are also thought to reside in halos that are susceptible to heating from the UV background, which cuts off gas accretion to the galaxy, removing fuel for star formation \citep[e.g.,][]{Brown2014,Weisz2014,Onorbe2015,Wheeler2015}.  Taken together, these processes suggest that UFDs cannot self-regulate.  Thus, the choice of prescriptions for star formation and feedback in cosmological simulations may strongly affect their resulting stellar mass, unlike in more massive galaxies.
In this paper we isolate the effect of star formation prescriptions that vary across simulations both in terms of the details of the simulation and the consequences for observational predictions.  


One of the key differences in star formation recipes between cosmological galaxy simulations is the threshold mass density at which star formation is allowed to occur.  By definition, lower  resolution simulations do not have the ability to resolve high density gas peaks.  Thus, the density threshold must vary with resolution of the simulation.  Star formation density thresholds vary from $\sim$1 $m_H$ cm$^{-3}$, where $m_H$ is the mass of a hydrogen atom  in grams, in lower resolution simulations \citep[e.g., APOSTLE,][]{apostle} to 10 $m_H$ cm$^{-3}$ \citep[e.g., NIHAO,][]{NIHAO} to $> 100$ $m_H$ cm$^{-3}$ \citep[e.g.,][]{G10, Shen2014, FIRE, FIRE2, Read2016}.  A maximum temperature threshold for star formation is also usually applied.  Many simulations adopt a temperature cap of $\sim$10$^4$ K because this is the peak of the cooling curve and gas is expected to rapidly cool to lower temperatures \citep[e.g., see][]{saitoh08}.  

Beyond this temperature-density model, some simulators also track the presence of molecular hydrogen, H$_2$, requiring that it be present in order to form stars.  The H$_2$-based star formation models broadly break down into two categories, equilibrium \citep{Krumholz2008,Krumholz2009,Kuhlen2012,FIRE,FIRE2} or non-equilibrium \citep{Robertson2008, Gnedin2009,Gnedin2010,Gnedin2011,Christensen2012}. Both models, by requiring the presence of H$_2$, ensure that stars form from high density gas: generally star formation occurs in gas with $n > 100$ $m_H$ cm$^{-3}$ but it can be as high as $n > 1000$ $m_H$ cm$^{-3}$ depending on metallicity and resolution. 
The temperature cap for star formation is also usually lowered in these models as additional cooling processes are captured.  The equilibrium models do not explicitly track the formation and destruction of H$_2$, but rather assume a two-phase interstellar medium in which formation and destruction are balanced. In the non-equilibrium model, the formation and destruction of H$_2$ are instead explicitly followed.  Thus, in the non-equilibrium models, star formation is dependent on the timescale of H$_2$ formation, which is not the case in the equilibrium models.   

Using an equilibrium H$_2$ model, \citet{Kuhlen2012} showed that an H$_2$-based star formation prescription could dramatically suppress star formation in low mass halos where metallicities are low (and thus H$_2$ is unable to form).  However, the suppression is weakened in H$_2$ formation models if dense gas is able to shield and form H$_2$ despite low metallicities when sufficiently high resolutions are achieved \citep[e.g.,][]{Hopkins2013c}. 
 In this paper, we show that adopting a non-equilibrium model leads to further changes.  At low metallicities, there is a delay in H$_2$ formation times in non-equilibrium models, leading to a quantitative difference in the ability of UFD galaxies to form stars compared to temperature-density threshold models. 

The two star formation prescriptions that we explore in this paper were also adopted in \citet[][``Metals'' and ``H$_2$'' in that work]{Christensen2014b}, where it was shown that both models produce dwarf galaxies with nearly identical structural parameters (rotation curves, dark matter density profiles, baryonic angular momentum distributions) for galaxies with $\sim$10$^8$ M$_{\odot}$ in stellar mass.  Only the mass in galactic winds varied, but by less than a factor of two.
We show in this work that, although these star formation prescriptions lead to similar galaxies in the classical dwarf galaxy mass range, differences arise on the scale of UFDs.

The differences that star formation prescriptions introduce on UFD scales have important ramifications, e.g., for the slope and scatter at the faint-end of the Stellar Mass -- Halo Mass (SMHM) relation and the slope of the stellar mass function in the ultra-faint regime \citep{Lin2016,Munshi2017}.  In this paper we also emphasize the impact on the expected number of UFD satellites in dwarf galaxies,  as this has been an active area of investigation recently \citep{Wheeler2015,Dooley2017a}, particularly with respect to possible companions of the Magellanic Clouds \citep{Dooley2017b,Deason2015, Yozin2015, Jethwa2016, Sales2017, Kallivayalil2018, Li2018}.
The favored $\Lambda$ Cold Dark Matter ($\Lambda$CDM) cosmological model predicts that structure formation is self-similar.  All dark matter halos should contain an abundance of dark matter substructure, from the largest galaxy cluster halos, to the smallest halos containing observed dwarf galaxies, and beyond.    
The scale-free nature of the subhalo mass function in $\Lambda$CDM suggests that groups of subhalos should be common \citep{Li2008,Donghia2008, Sales2011, Nichols2011}. 
Because low-mass halos form earlier, are denser, and fall into smaller hosts before larger ones, it is likely that satellites of satellites or of low mass isolated halos have survived longer than their counterparts that fell directly into the Milky Way \citep{Diemand2005}. This suggests that one way to search for ultra-faint galaxies might be to search for satellites of known dwarf galaxies \citep{Rashkov2012,Sales2013,Wheeler2015,madcash,Patel2018}. 
%
%
%
\citet{Dooley2017a} used a range of SMHM relations derived from abundance matching results combined with a model for reionization to show that isolated dwarfs in the Local Group are extremely interesting targets in the hunt for ever-fainter dwarfs. 

This paper is organized as follows: we describe our simulations in Section \ref{Sims}. In Section \ref{Results} we quantify the occupation fraction of dark matter halos as a function of declining halo mass, and show that at the lowest halo masses there is a drastic difference in the number of luminous dwarf galaxy satellites in simulations with different star formation prescriptions.  We find that the inability of low mass halos to form H$_2$ in the reionization epoch suppresses star formation relative to the same halos run with a temperature-density threshold star formation model.   In Section \ref{Discussion}, we demonstrate how the choice of star formation implementation impacts various quantities (star formation histories, stellar mass function, and probability of a classical dwarf hosting UFD satellites) that have recently been studied using simulations of dwarf galaxies.
We discuss our results, including limitations of our model, in Section \ref{discuss}.
We summarize in Section \ref{Summary}.

\section{The Simulations}\label{Sims}

The simulations used in this work are run with the N-Body + Smoothed Particle Hydrodynamics (SPH) code {\sc ChaNGa} \citep{Menon2015} in a fully cosmological $\Lambda$CDM context using WMAP Year 3 cosmology: $\Omega_0=0.26$, $\Lambda$=0.74, $h=0.73$, $\sigma_8$=0.77, n=0.96. 
{\sc ChaNGa} utilizes the {\sc charm}++ run-time system for dynamic load balancing and computation/communication overlap in order to effectively scale up the number of cores.  This improved scaling has allowed for the simulation of large, high-resolution volumes \citep[e.g.,][]{Tremmel2017, Anderson2017} that were previously unattainable with {\sc ChaNGa}'s predecessor code, {\sc Gasoline}.  We use this scaling here to simulate a volume of dwarf galaxies,  twice, requiring a total of $\sim$10 million CPU hours.

{\sc ChaNGa} adopts all the same physics modules (described below) as {\sc Gasoline}, but has an improved SPH implementation that uses the geometric mean density to more realistically model the gas physics at the hot-cold interface \citep{Wadsley2017}.  The developers of {\sc ChaNGa} are part of the {\sc agora} collaboration, which aims to compare the implementation of hydrodynamics across cosmological codes \citep{Kim2014, Kim2016}.

Our galaxy sample is drawn from a uniform dark matter-only simulation of a 25 Mpc per side cube.  From this volume, we select a field--like region representing a cosmological ``sheet'' roughly 3 Mpc in diameter, containing almost 7000 isolated dark matter halos from 2 $\times$10$^{10}$ M$_{\odot}$ in halo mass down to our resolution limit of 4.3 $\times$ 10$^5$ M$_{\odot}$ (64 particles). 
We then re-simulate this field at extremely high resolution using the ``zoom-in'' volume renormalization technique \citep{katz93}. The zoom-in technique allows for high resolution in the region of interest, while accurately capturing the tidal torques from large scale structure that deliver angular momentum to galaxy halos \citep{barnes87}. These zoom-in simulations have a hydrodynamical smoothing length as small as 6 pc, a gravitational force softening of 60 pc, and an equivalent resolution to a 4096$^3$ particle grid.  Dark matter particles have a mass of 6650 M$_{\odot}$, while gas particles begin with a mass of 1410 M$_{\odot}$, and star particles are born with 30\% of their parent gas particle mass. 
The dark matter-only version of this volume was run in both a CDM and self-interacting dark matter (SIDM) model in \citet{Fry2015}. Following the convention in that paper, we adopt the nickname ``The 40 Thieves.'' 

\medskip
\noindent {\bf Metal Cooling (MC):} 
 The ``Metal Cooling (MC)'' version of the 40 Thieves includes cooling of the gas via primordial and metal-line cooling, non-equilibrium abundances of H and He, and diffusion of metals to neighboring gas particles \citep{shen10}. 
Additionally, we adopt a simple model for self-shielding of the HI gas following \cite{pontzen08}.  
Star formation occurs stochastically when gas particles become cold (T$< 10^4$K) and when gas reaches a density threshold of 100 $m_H$ $\rm cm^{-3}$, comparable to the mean density of molecular clouds.    
The probability, $p$, of spawning a star particle is a function of the local dynamical time $t_{form}$:
\begin{equation}
p = \frac{m_{gas}}{m_{star}} \left (1 - e^{-c^* \Delta t/t_{form}} \right)
\label{eqn1}
\end{equation}
where $m_{gas}$ is the mass of the gas particle and $m_{star}$ is the initial mass of the potential star particle. A star formation efficiency parameter, $c^* = 0.1$, gives the correct normalization of the Kennicutt-Schmidt relation \citep{Christensen2014b}.

\medskip
\noindent {\bf Molecular Hydrogen (H$_2$):} The ``Molecular Hydrogen (H$_{2}$)'' version of the 40 Thieves includes the aforementioned metal line cooling and metal diffusion, with the addition of non-equilibrium H$_2$ abundances and H$_2$-based star formation.
Our H$_2$ abundance calculation includes a both a gas-phase and dust-dependent description of H$_2$ creation and destruction.  H$_2$ forms via H$^-$ in the gas phase, and is collisionally dissociated. The dust phase is dominant as soon as a small amount of metals are present. A detailed discussion of the calculation of H$_2$ formation on dust grains is given in \citet[hereafter CH12]{Christensen2012}. Destruction of H$_2$ by Lyman-Werner (LW) radiation is calculated due to both the UV background and from nearby young stars.  An extensive calculation of the photodissociation by LW radiation is found in CH12. 
Shielding of HI and H$_2$ is based on particle metallicity 
\citep[CH12]{gnedin09}.  As in CH12, the SFR in this simulation is set by the local gas density and the H$_2$ fraction.  The star formation probability is again given by equation \ref{eqn1}, but $c^*$ is modified such that $c^* = c{_0^*} X_{H_2}$, 
where $c{_0^*} = 0.1$ and $ X_{H_2}$ is the fraction of baryons in $H_2$.
We restrict star formation to occur in gas particles with T $< 10^3$ K.  With the inclusion of the H$_2$ fraction term, gas in low-metallicity dwarfs tends to reach even higher densities (required for the gas to shield and form H$_2$) than the MC model before it can form stars \citep{Christensen2014b}.

\medskip
We apply a uniform, time-dependent UV field from \cite{HM2012} to model photoionization and photoheating for both runs. Both simulations 
adopt the ``blastwave'' supernova feedback approach \citep{stinson06}, in which mass, thermal energy, and metals
are deposited into nearby gas when massive stars evolve into supernovae.  The thermal energy deposited amongst those nearby gas neighbors is 10$^{51}$ ergs per supernova event. Following \citet{Stinson2010}, we quantize the feedback so that supernovae only occur when whole stars have gone supernova, as opposed to slowly releasing fractions of supernova energy at every time step. Subsequently, gas cooling is turned off until the end of the
momentum-conserving phase of the supernova blastwave. 
This model keeps gas hydrodynamically coupled at all times.  

\begin{figure*}
\centering
\includegraphics[width=1.9\columnwidth]{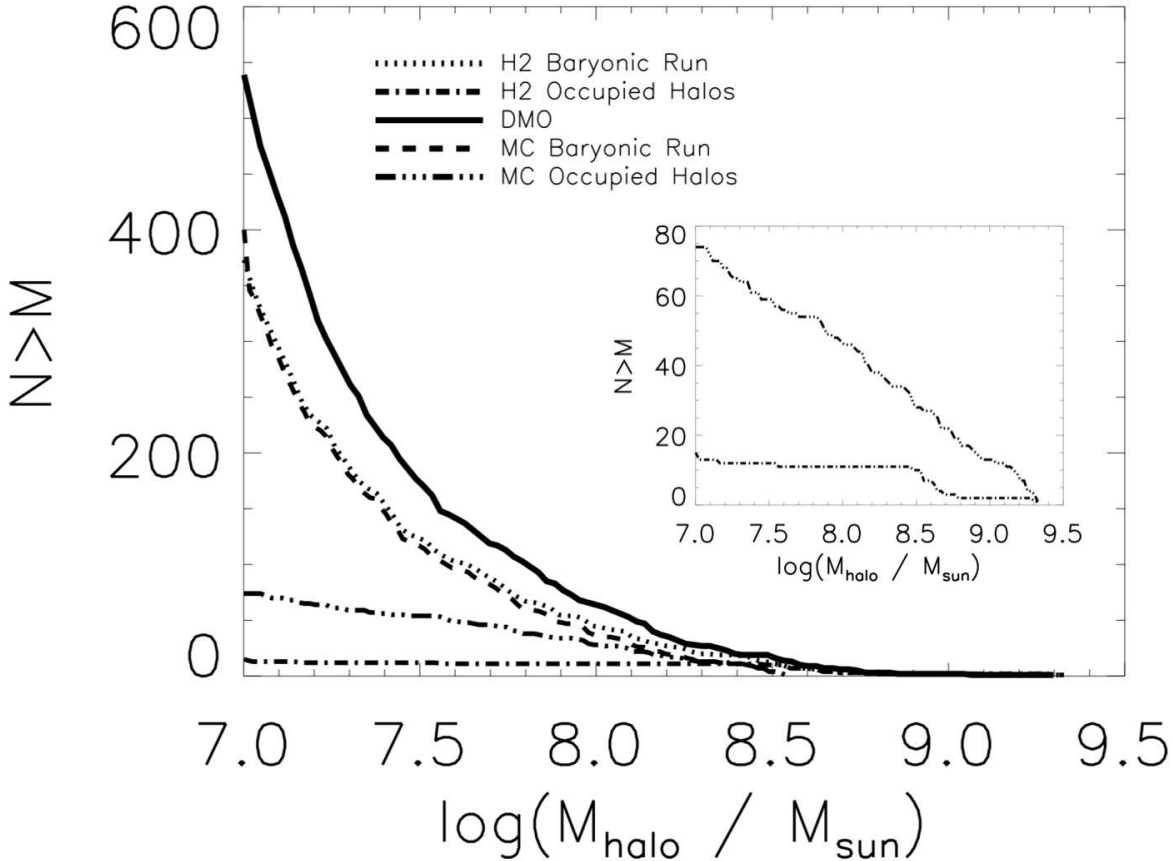}
\caption{Cumulative halo mass functions. 
 The solid line is the cumulative mass function for the dark matter-only (DMO) run; the dashed and dotted lines are the cumulative mass function for the two baryonic runs using all halos; the two dot-dashed lines are the cumulative mass function using only occupied (luminous) halos in the baryonic runs. Inset: A zoom in of the occupied halos in both the H$_2$ and MC runs.  
The fraction of dark (non-luminous) halos continues to increase with decreasing halo mass in the MC run, but remains constant in the H$_{2}$ run below $\sim$ $10^{8.5}$ M$_{\odot}$ in halo mass.  The H$_2$ run contains far fewer occupied halos than the corresponding MC run, by roughly a factor of five. 
}
\label{fig:dark}
\end{figure*}

Our feedback model does not explicitly include processes such as cosmic rays, or those caused by young stars such as photoionization, momentum injection from stellar winds, and radiation pressure \citep[e.g.,][]{Thompson2005, Wise2012,Murray2011,Hopkins2012, Sharma2012,Agertz2013, Booth2013, Simpson2016, Salem2016,Farber2018,Kannan2018}.
However, the physical prescriptions described above have been able to reproduce and explain properties of galaxies over a wide range of masses, regardless of which SF recipe is adopted.  In addition to simulating the first bulgeless disk galaxy and dark matter cores \citep{G10,brook11}, simulated galaxies match the observed mass -- metallicity relation \citep{brooks07, Christensen2016}, the baryonic Tully-Fisher relation \citep{Christensen2016, Brooks2017}, the size -- luminosity relation \citep{brooks11}, the 
stellar mass to halo mass relation determined from abundance matching \citep{Munshi13}, and the sizes and fractions of HI in local galaxies \citep{Brooks2017}.  They also match the abundance of DLA systems \citep{pontzen08}, and the numbers and internal velocities of dwarf Spheroidal satellites \citep{BZ2014}. 
In what follows, we extend these successful models to lower masses and demonstrate for the first time that the differing star formation models impact galaxy formation on UFD scales.

Halos are identified and tracked with Amiga's Halo Finder \citep[AHF][]{gill04,knollmann09}. AHF calculates the virial mass of each halo (given in this paper by M$_{halo}$) as the total mass with a sphere that encloses an overdensity relative to the critical density of 200$\rho_{crit}(z)$.

 \section{Results}\label{Results}
 
Figure \ref{fig:dark} shows the cumulative halo mass functions for both of the baryonic (MC and H$_2$) and dark matter-only versions of the 40 Thieves.  The top three lines (solid, dashed, and dotted) include all dark matter halos (both isolated and satellite galaxies) down to 10$^7$ M$_{\odot}$ in halo mass (corresponding roughly to the hydrogen cooling limit) at $z=0$, regardless of whether they contain stars.  The bottom two lines are only those halos that are ``occupied'' by stars, and are also shown separately in an inset for clarity.  
A given dark matter halo in a baryonic run is less massive than in the corresponding dark matter-only simulation \citep[see also][]{Sawala2013,Munshi13}.
The root cause of this mismatch is baryon ejection from low mass halos (either by heating from the UV background and/or as a result of supernova feedback) which slows not only the galaxy growth rate, but the dark matter halo growth rate as well.  
As a direct result, the total number of dark matter halos with M$_{halo} > $10$^7$ M$_\odot$ is reduced ($\sim$75\%) in the baryonic versions compared to the dark matter-only run. 

\begin{figure}
\centering
\includegraphics[width=1\columnwidth]{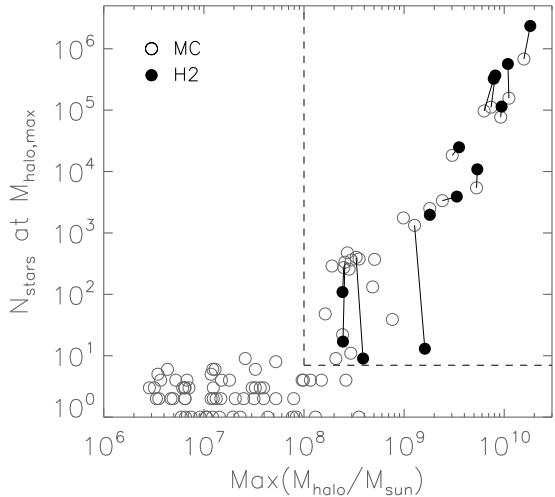}
\caption{The number of stars in a halo at the time the maximum halo mass is attained versus the maximum halo mass, for both the MC and H$_2$ runs.  The dashed line separates the galaxies in our matched sample (M$_{halo} > 10^8$ M$_{\odot}$ and $\geq$ 7 star particles). Lines connect matched galaxies in both the MC and H$_2$ runs.  Above $10^9$ M$_{\odot}$ in halo masses, the galaxies are well-matched across runs, but below $10^9$ M$_{\odot}$ there are more galaxies produced in the MC run than the H$_2$ run. }
\label{fig:converge}
\end{figure}

The inset of Figure \ref{fig:dark} shows in closer detail the cumulative halo mass function for only those halos that ``host a galaxy,'' i.e., contain a minimum of one star particle \citep[see also][]{Sawala2015} in both the MC and H$_{2}$ runs. 
The number of luminous halos continues to rise toward lower halo masses in the MC run, but stops rising below M$_{halo}$ $\sim$ $10^{8.5}$ M$_{\odot}$ in the H$_{2}$ run.
There are nearly five times as many occupied halos above $10^7$ M$_\odot$ in the MC run than in the H$_{2}$ run.  
If we focus only on the higher mass halos in our matched sample 
(defined next), the difference drops to about a factor of two (see Figures \ref{fig:converge} and \ref{fig:hugeplot}).

\subsection{Matched Sample}

 The galaxies with one star particles are, by definition, not resolved.  In this section we set out to identify a set of halos that can be reliably compared against each other from each run. For every dark matter halo that contains a star particle at $z=0$, we trace back the most massive progenitor halo and identify the time step in which it has the maximum number of dark matter particles.  In Figure \ref{fig:converge} we show the halo mass and number of star particles inside the halo at that step.  
Solid lines connect matched galaxies across the MC and H$_2$ versions of the simulation. For halos above $10^9$ M$_{\odot}$, halos are well-matched, with a one-to-one correspondence between formed galaxies.  Below $10^9$ M$_{\odot}$, however, it is clear that the MC run forms more galaxies than the H$_2$ run, and even in the matched cases the H$_2$ galaxies tend to form fewer stars at these low halo masses.

For halos with maximum halo mass above $10^8$ M$_{\odot}$ and N$_{star} \ge 7$,  there are MC halos that contain galaxies but with matched counterparts in the H$_2$ run that are completely dark.  We have examined those dark halos in more detail to test whether their lack of star formation is expected for the H$_2$ model given the resolution.  We have approached this in two ways.  
First, similar to \citet{Kuhlen2013}, we compare the surface densities of our gas particles to the critical surface density $\Sigma_{crit}$ at which atomic H converts to molecular H.  The metallicity of our non-star forming gas is $Z/Z_{\odot} = 10^{-3}$ or lower.  At $Z/Z_{\odot} = 10^{-3}$, $\Sigma_{crit} = 5700$ M$_{\odot}$ pc$^{-2}$. The non-star forming gas in the dark matched H$_2$ halos remains at surface densities $< 10$ M$_{\odot}$ pc$^{-2}$.  Thus, it would not be capable of forming H$_2$, and therefore should not form stars in the H$_2$ run \citep[see also][]{Sternberg2014}.

Second, we can examine the timescale, $t_{H_2}$, for atomic H to convert to molecular H following \citet{Krumholz2012}.  Following their equation 7:
\begin{equation}
\frac{t_{H_2}}{t_{ff}} = 24 Z^{-1} C^{-1} n_0^{-1/2}
\label{eq2}
\end{equation}
where $t_{ff}$ is the free-fall time, $Z$ is the metallicity of the gas relative to solar, $C$ is a clumping factor equal to 10 in our simulations, and $n_0 = \avg {n_H}/1$ cm$^{-3}$ where $\avg {n_H}$ is the mean hydrogen density.  For the non-star forming gas in our matched H$_2$ halos, $\avg {n_H}$ is about 10 $m_H$ cm$^{-3}$.  The free-fall time depends on $\avg {n_H}$ as well, and is 13.6 Myr at 10 $m_H$ cm$^{-3}$.  The timescale for H$_2$ formation, $t_{H_2}$, is more than 10 Gyr for our gas with $Z/Z_{\odot} = 10^{-3}$.  Even at higher densities of $\avg {n_H} = 100$ $m_H$ cm$^{-3}$, the timescale for H$_2$ formation is over 1 Gyr.  

Thus, significant amounts of H$_2$ simply cannot form in these halos before reionization  at this resolution (we discuss limitations of the resolution in Section \ref{discuss}).  We conclude that our dark halos in the H$_2$ run are behaving as expected. 
We therefore restrict the following analysis to matched halos with M$_{halo,max} > 10^8$ M$_{\odot}$ and N$_{stars} \ge 7$. It should be clear that we do not mean that these halos are ``converged'' across the two star formation recipes.  The star formation in the two runs in the lowest mass halos is dramatically different, with the MC run forming stars while matched halos in the H$_2$ run remain dark.  Even when the H$_2$ run forms stars there is a discrepancy in stellar mass with the matched halos in the MC run for halos with M$_{halo,max} \lesssim 10^9$ M$_{\odot}$.  This is exactly the difference we wish to explore in this work.  %
 
\subsection{The Formation of Dwarf Galaxies}

\begin{figure*}
\centering
\includegraphics[width=2.0\columnwidth]{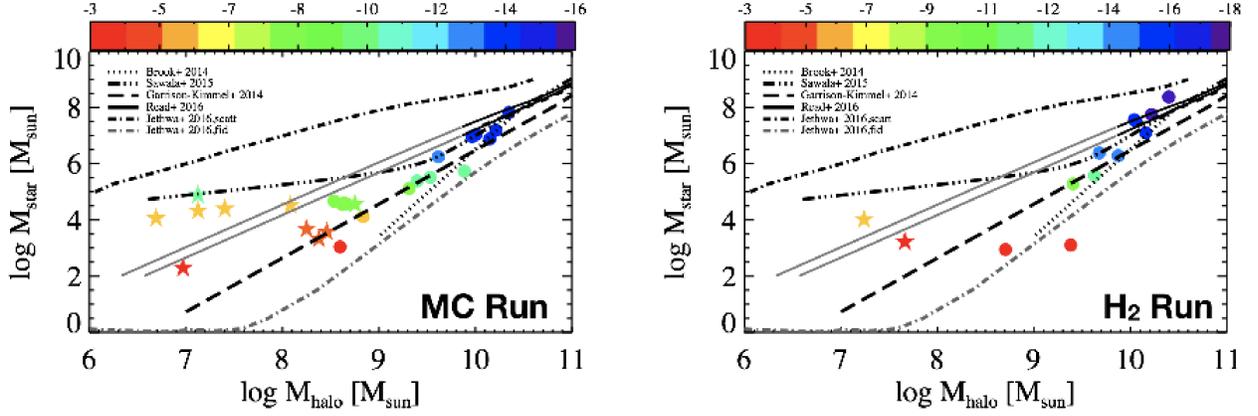}
\caption{Stellar to Halo Mass relationship for both simulations, showing only the matched galaxies from Figure \ref{fig:converge}. {\it Left panel}: MC Run. {\it Right panel}: H$_{2}$ run. Circles are central galaxies and stars are resolved satellites.
Galaxies are color-coded by their $V$-band magnitude (see color bar on top).  Stellar masses are derived based on photometric colors, and halo masses are taken from the DM-only version of the run to be consistent with the abundance matching results shown \citep{Brook2014,GarrisonKimmel2014,Read2017, Jethwa2018}. We plot $z=0$ halo masses. We also show simulation results from \citet{Sawala2015}. A similar number of central galaxies form in both simulations for M$_{halo}$ $>$ 10$^9$ M$_{\odot}$, but more galaxies form in the MC run at lower halo masses, and there are many more luminous satellites in the MC run.}
\label{fig:hugeplot}
\end{figure*}

In Figure \ref{fig:hugeplot} we show the resulting stellar mass to halo mass relation for matched halos in both the MC (left panel) and H$_2$ (right panel) versions of the 40 Thieves.  Each galaxy is color coded by its $V$-band magnitude at $z=0$.  Filled circles are central (isolated) galaxies, while stars are satellite galaxies. Four of the lines show results from abundance matching in dwarf galaxies \citep{Brook2014,GarrisonKimmel2014,Read2017, Jethwa2018}, and the fifth line shows the simulation results of \citet{Sawala2015}. 
The stellar masses have been calculated following \citet{Munshi13}, based on photometric colors.  To be consistent with abundance matching, the halo masses are the mass in the corresponding dark matter-only run.  Note that we plot $z=0$ halo masses, while the abundance matching results use maximum halo mass.  Thus, the satellite results should not be compared directly to the lines (though we note that all satellites must have peak halo masses above 10$^8$ M$_{\odot}$ to be included in the sample shown here).

Again, it is immediately obvious from Figure \ref{fig:hugeplot} that there are galaxies residing in halos above M$_{halo}$ $\sim$10$^9$ M$_{\odot}$ that are produced in both runs.  However, below $\sim$10$^9$ M$_{\odot}$ the number of galaxies diverges, with twice as many halos containing galaxies in the MC run.  We can now see from Figure \ref{fig:hugeplot} that many of these low mass halos are satellites (colored stars).  There are five times as many satellites in the MC run than in the H$_2$ run.

\begin{figure}
\centering
\includegraphics[width=1.\columnwidth]{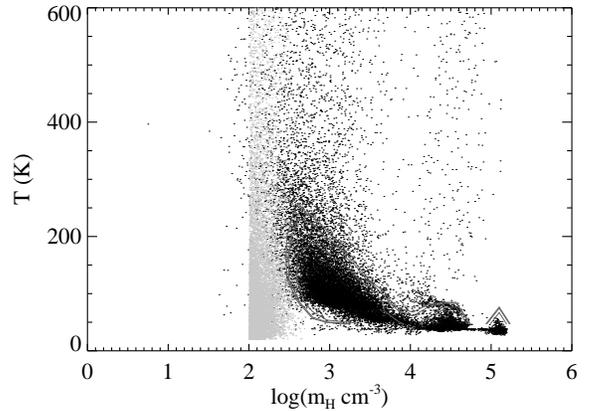}
\caption{Comparison of the phase diagrams for star forming gas between the MC run (grey points) and the H$_2$ run (contours + black points) for all stars formed in the first Gyr of the simulation.  Note that the range of densities forming stars in the H$_2$ run, while slightly overlapping with that in the MC run, overwhelmingly tend to be higher than those in the MC run.}
\label{fig:phase}
\end{figure}

\begin{figure*}
\centering
\includegraphics[width=1.0\columnwidth]{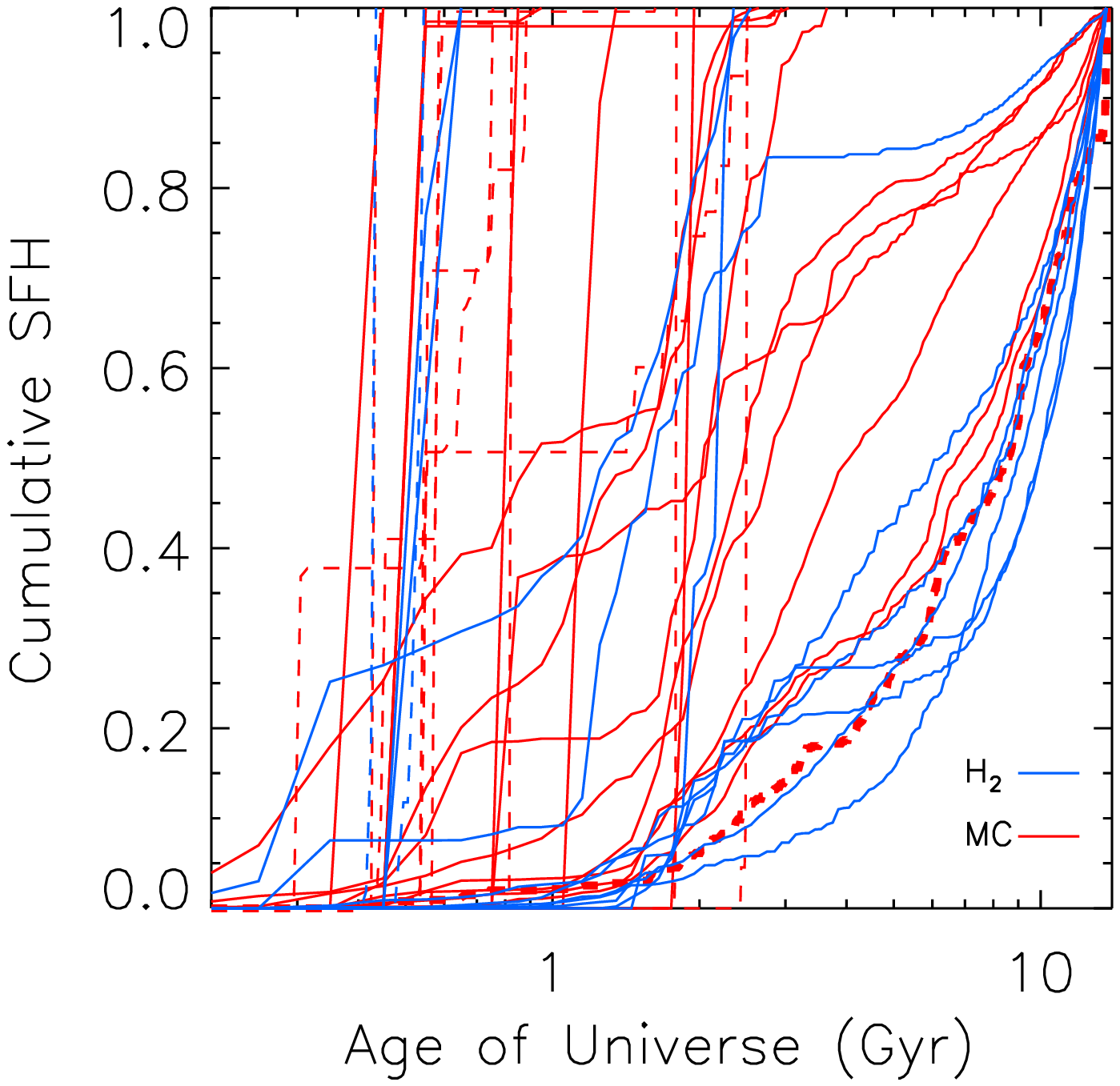}\includegraphics[width=1.0\columnwidth]{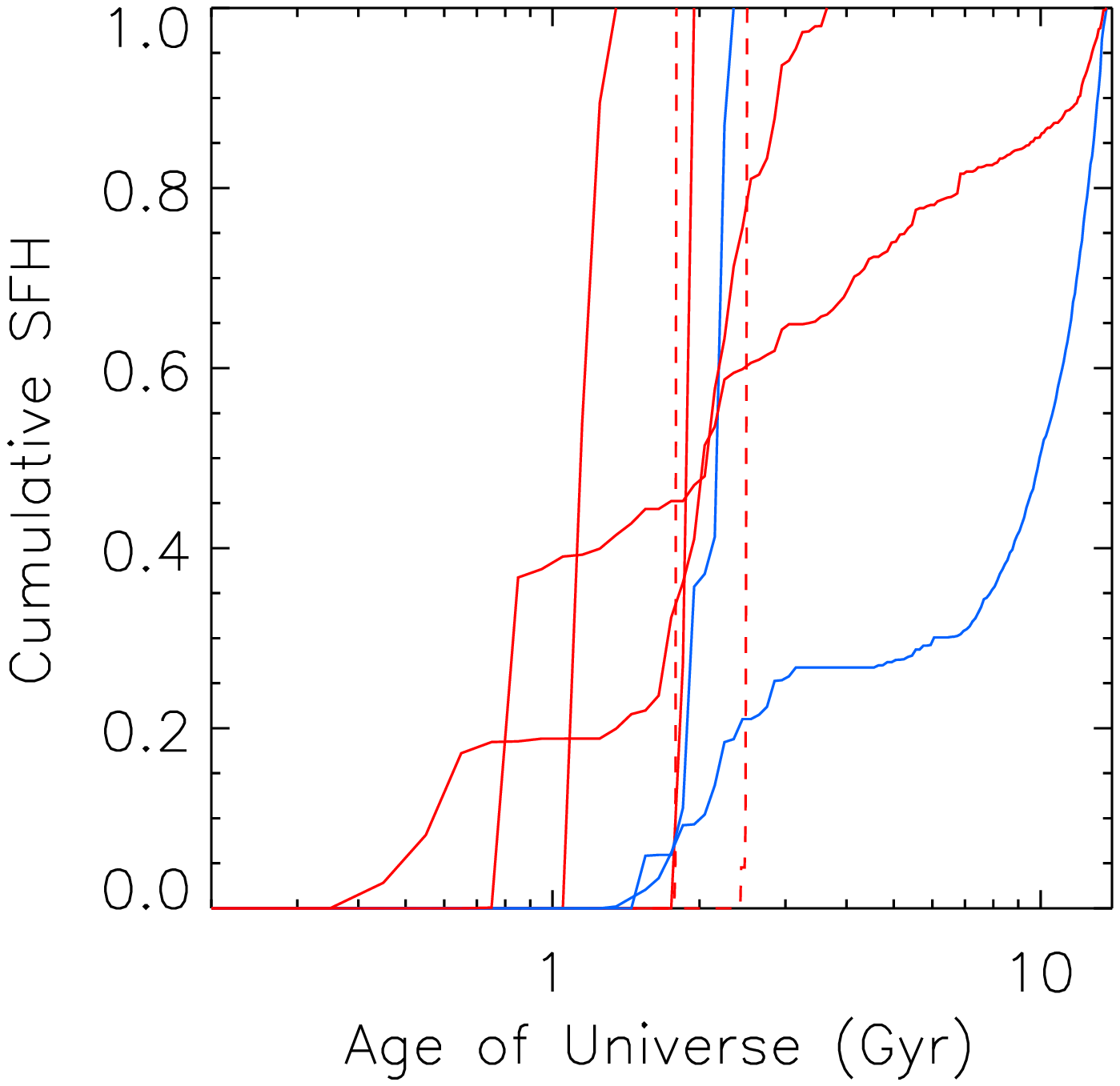}
\caption{Cumulative SFHs for the simulated galaxies. The MC run is shown in red, the H$_2$ run in blue.  Solid lines are central galaxies at $z=0$, while dashed lines are satellites.  {\it Left Panel}: All surviving matched galaxies at $z=0$. One MC satellite is shown in a thicker dashed line that has an extended SFH, but its H$_2$ counterpart has been destroyed in a merger, as discussed in Section \ref{delaymerge}. {\it Right Panel}: A subset of the galaxies shown in the left panel that are discussed in Sections \ref{delaymerge} and \ref{delaysf}.  The four MC galaxies that do not begin forming stars until after 1 Gyr are shown (two satellites -- red dashed, two centrals -- solid red).  The two H$_2$ galaxies that don't start star formation until after 1 Gyr are shown (solid blue) along with their MC counterparts that start forming stars much earlier (solid red). }
\label{fig:sfhs}
\end{figure*}

All of the satellites in these runs are in the ultra-faint luminosity range. 
Like observed UFDs \citep{Brown2012,Brown2014,Weisz2014,Weisz2014b}, they tend to form the bulk of their stars early, with their star formation trickling off soon after reionization (see Figure \ref{fig:sfhs}, discussed further below). 
In Figure \ref{fig:phase} we compare the phase diagram for gas that forms stars within the first 1 Gyr of both simulations (i.e., before the end of reionization at $z \sim 6$).  Although there is some overlap in the star formation temperatures and densities between the two runs, there is a clear offset in the regions where the majority of stars form.  In the MC run (grey points), stars tend to form from gas that spans a range of temperatures (up to 10$^4$ K) but is near the threshold density (100 $m_H$ cm$^{-3}$).  In the H$_2$ run (contours and black points), star-forming gas forms from colder, denser gas ($< 500$ K and $\gtrsim$ 1000 $m_H$ cm$^{-3}$).  This difference was also discussed in CH12 (see their Figure 13).

The higher densities that stars form from in the H$_2$ model is tied directly to the subgrid H$_2$ formation model. At solar metallicities, the two models form star particles at similar densities, but the models diverge as metallicity decreases. As described in CH12, when metals are present the formation of H$_2$ is dominated by formation on dust grains. In the reionization epoch when metallicities are extremely low, formation may also proceed through gas-phase reactions.  However, during the reionizaton epoch both of these processes are inefficient and slow. 
As a result, gas particles will frequently reach high densities through gravitational collapse prior to forming significant amounts of H$_2$. At the same time, the low metallicities reduce the amount of dust-shielding, which means that H$_2$ (and thus star formation) can only persist in high density gas. For most of the H$_2$ subhalos, significant amounts of H$_2$ can never be created before reionization removes the gas from the halos.  In the MC run, because stars can form in lower density atomic gas, star formation can begin before reionization quenches it.  This also explains why the MC run tends to form more stars in general for galaxies in halos with M$_{halo}$ $<$ 10$^9$ M$_{\odot}$ (see Figure \ref{fig:converge}).


Even at $z=0$, the most massive matched dwarfs in this work still show a difference in the effective star formation density threshold due to the impact of metallicity on dust-shielding \citep[CH12;][]{Christensen2014b}.  Despite this, dwarf galaxies in halos above $\sim$10$^9$ M$_{\odot}$ generally still form similar numbers of star particles, due to their ability to self-regulate.  This is consistent with the prior results discussed in the Introduction that found that the density threshold had little impact on the resulting SFR. However, a slight excess of stars seems to form in the H$_2$ model relative to the MC model at these masses (see Figure \ref{fig:converge}). 
This excess is likely due to the ability of the gas to shield in the H$_2$ model, protecting the gas from heating and leading to additional cold gas present in the H$_2$ simulations. 
The excess cold gas results in slightly higher stellar masses in the H$_2$ run in halos $\sim$10$^{10}$ M$_{\odot}$.

\citet{Kuhlen2012} were the first to show that a model for H$_2$-based star formation could suppress star formation in dwarf galaxies primarily due to their lower metallicities. 
However, they found that their equilibrium H$_2$ model suppressed all star formation in halos below about M$_{halo}$ $=$ 10$^{10}$ M$_{\odot}$, while our non-equilibrium H$_2$ model forms stars in halos down to M$_{halo}$ $=$ 10$^{8.5}$.  
Reionization played no role in \citet{Kuhlen2012}, since their lowest mass halo that was able to form stars was well above the halo mass thought to be impacted by reionization.  On the other hand, the changes in stellar mass that we see in halos below $\sim$10$^9$ M$_{\odot}$ are explicitly due to the fact that these are halos in which reionization can strongly affect the evolution. The MC halos are able to start forming stars at lower densities than their H$_2$ counterparts, and thus are able to produce more stars before reionization removes their gas.




\subsection{Delayed Merging}\label{delaymerge}

Early star formation is not the only reason why there are more satellites in the MC run than in the H$_2$ run: more of the satellites survive.  
Three of the surviving subhalos in the MC run are completely disrupted in the H$_2$ run after merging with a parent halo. Two of those halos had managed to form stars in the H$_2$ run before being fully destroyed.  In fact, a close examination of Figure \ref{fig:sfhs} shows that there is a surviving satellite in the MC run with a very extended star formation history (SFH; bolded red dashed line in the left panel).  This satellite had a counterpart in the H$_2$ run with an extended SFH, but the subhalo is completely disrupted and part of the second largest halo in the H$_2$ run at $z=0$.  All three of the surviving subhalos in the MC run are completely merged in the dark matter-only run as well.

\citet{Schewtschenko2011} showed that mergers occur later in simulations with baryonic feedback than in matched halos in an identical dark matter-only simulation.  They hypothesized that the pressure of the hot halos found in baryonic simulations slows down accretion of subhalos.  Our simulated dwarf galaxies {\it do} have hot halos of gas around them \citep{Wright2018}, with the mass in hot gas being up to an order of magnitude greater than the HI gas mass. Later infall of subhalos then leads to less tidal mass loss for subhalos simply due to the fact that there is less time for tidal stripping between infall and $z=0$ (Ahmed et al., in prep.). The end result should be that fewer satellites are fully destroyed at $z=0$ for a run with baryonic feedback. \footnote{This assumes that additional tidal effects from the potential of the central galaxy are negligible.  The disk potential is not negligible in Milky Way-mass galaxies \citep{Penarrubia2010,Zolotov2012,Arraki2014,BZ2014,GarrisonKimmel2017}, but is expected to be negligible in dwarf galaxies.}

These results suggest that the feedback in the MC run is stronger, delaying mergers of subhalos.  Indeed, another indication that feedback is stronger in the MC run is that the overall halo masses tend to be lower (see comparison of maximum M$_{halo}$  for matched halos in Figure \ref{fig:converge}).  As mentioned previously, feedback reduces the growth of halos \citep{Munshi13,Sawala2013}.  Despite the fact that the feedback recipe is the same in both the MC and H$_2$ runs, 
feedback is injected into lower density gas in the MC run because the MC run rarely reaches the higher densities found in the H$_2$ run \citep[see, e.g., figure 7 of][]{Christensen2014b}.
When comparing the nine most massive isolated dwarfs at $z=0$ (where the stellar masses are similar), the H$_2$ dwarf galaxies on average have 10\% higher dark matter mass and more than twice the mass in hot gas within their virial radii (and more baryons generally) compared to their MC counterparts.\footnote{Since the hot gas mass of the MC galaxies is smaller than the H$_2$ runs, the delayed merging of subhalos does not seem to be a direct result of pressure from the hot halo \citep[as was proposed in][]{Schewtschenko2011}, but rather a response to the ejection of the gas itself.} Thus, it seems that the MC galaxies have been able to remove more baryonic material from their halos, delaying their growth in dark matter as well.  

However, this result is seemingly at odds with previous comparisons of these two models \citep{Christensen2014}, which found that the H$_2$ model had more effective feedback and drove slightly more outflows.  While the previous work used the {\sc Gasoline} code, we use ChaNGa with an updated SPH treatment \citep{Wadsley2017} and quantized feedback.  It is not clear if these changes alter the feedback, making the MC run more effective at removing baryons.  A full investigation is beyond the scope of this paper, but we find that all evidence points to stronger feedback in the current MC runs than in the H$_2$ runs.


In summary, these results suggest that feedback can lead to delayed mergers of the UFD satellites. Combined with the ability to form stars at early times in more halos in the MC run, this makes the difference in satellite numbers even more stark at $z=0$ between the two runs.

\section{Implications}\label{Discussion}

The fact that star formation is restricted to different types of gas in these two commonly adopted models leads to some implications that should be considered when comparing predictions from different simulations.   Here, we examine a few observables that are commonly predicted by simulators, demonstrate that future observations have the potential to pinpoint more accurate physical models.

\subsection{Delayed Star Formation}\label{delaysf}

Consistent with the results presented in Section \ref{Results}, we find that there is a delay in the onset of star formation in the H$_2$ run compared to matched halos in the MC run.  We highlight a few cases of this in the right panel of Figure \ref{fig:sfhs}.  
In general (as can be seen in the left panel of Figure \ref{fig:sfhs}), most of our galaxies begin to form stars before $z=6$, i.e., in the first 1 Gyr after the Big Bang.
However, there are two (central) galaxies in the H$_2$ run that only begin to form stars after 1 Gyr (shown in blue).  Because their counterparts in the MC run can form stars without first generating significant amounts of H$_2$, the MC counterparts all begin star formation before the end of reionization (the counterparts are also shown in red in the right panel of Figure \ref{fig:sfhs} -- they are the two MC galaxies shown that begin star formation before 1 Gyr).

However, post-reionization onset of star formation cannot be attributed to only a single star formation model.  There are four galaxies (two centrals and two satellites, shown in the right panel of Figure \ref{fig:sfhs}) in the MC run that do not begin to form their stars until after 1 Gyr.  In each of the four cases, their counterparts in the H$_2$ run are dark halos that never formed stars.  In that case, we might say that the star formation is so delayed in the H$_2$ run that it does not occur before $z=0$.

\begin{figure*}
\centering
\includegraphics[width=1.5\columnwidth]{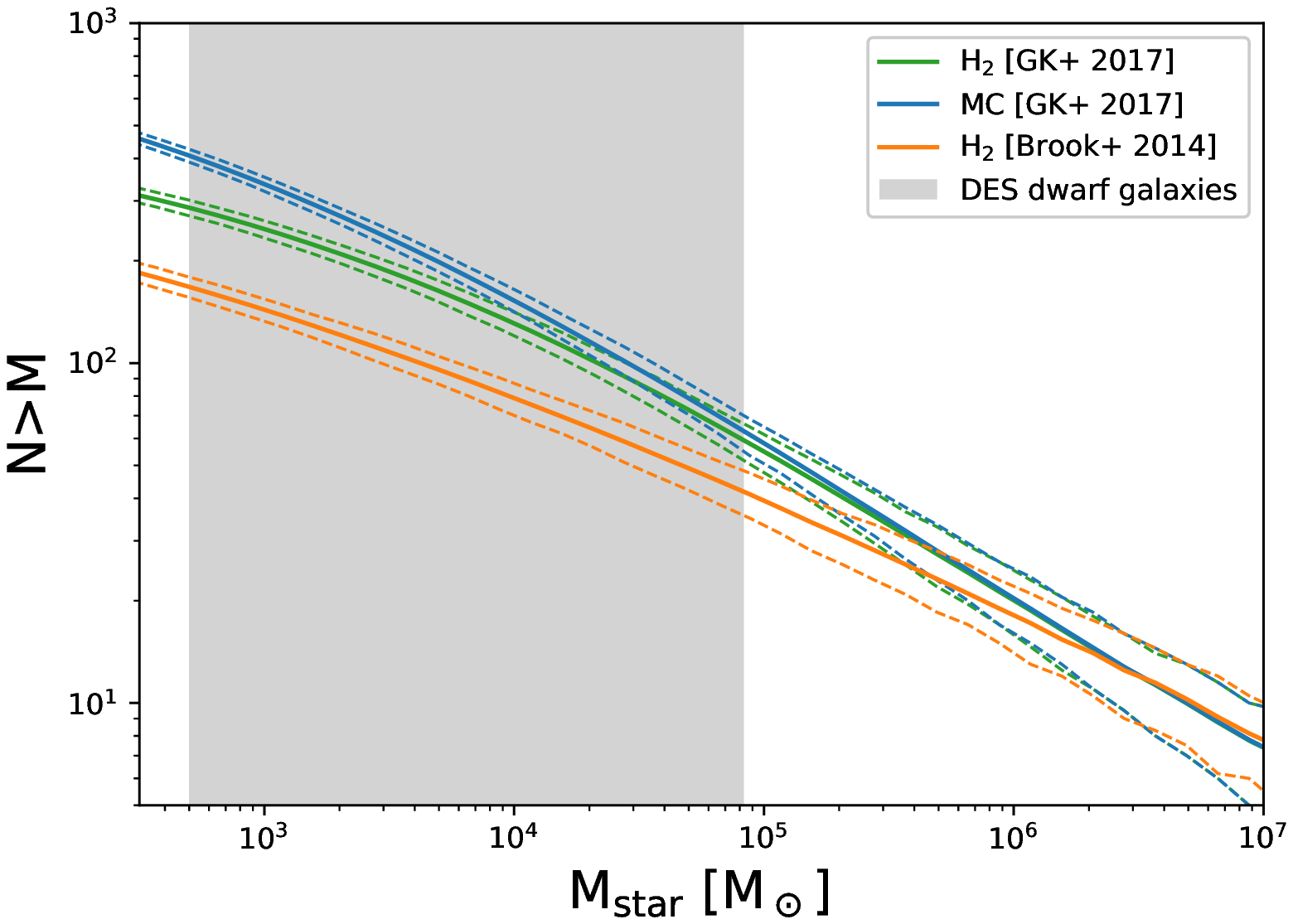}
\caption{Comparison of stellar mass functions predicted by the H$_2$ run (green) and the MC run (blue) assuming the slope and scatter of the SMHM relation follows \citet{GK2017} and following the occupation fractions shown in Figure \ref{fig:dark}.  The orange line is the SMF predicted for the H$_2$ run assuming the slope of the SMHM better follows \citet{Brook2014}. The grey shaded region represents the recent discovery space for DES. 1-$\sigma$ errors for each mass function are shown in corresponding colored dashed lines.}
\label{fig:mch2smf}
\end{figure*} 

In general, all observed galaxies have early star formation, though the error bars on the time of onset can be 1 Gyr or more depending on how far down the main-sequence resolved stars are detected \citep{Brown2012,Brown2014,Weisz2014,Weisz2014b}. It is not clear if our galaxies that delay star formation until after reionization are consistent with observations.  However, the dwarf galaxies in these simulated volumes are much further away from a massive galaxy than any observed galaxies with resolved star formation histories that have pushed below the oldest main-sequence turnoff.  Our dwarf galaxies are $\sim$5 Mpc away from a Milky Way-mass galaxy.  It remains to be seen if environment plays any role in onset of star formation. 
 
In summary, while the H$_2$ run has consistently later star formation (or none at all by $z=0$) compared to matched galaxies in the MC run, there is no obvious trend in onset time that could be used to discriminate models based on observations.  Rather, it is the number of galaxies and stellar mass of those galaxies that discriminates the models (discussed in the next section).

\subsection{Stellar Mass Function}

The factor of two difference in the number of faint dwarf galaxies that form between our two models will lead to different faint ends of the stellar mass function (SMF). 
Additionally, the fact that the MC run forms higher stellar masses for matched galaxies with M$_{halo,max} < 10^9$ M$_{\odot}$ (see Figure \ref{fig:converge}) will also alter the SMF.  
 
First, we demonstrate the difference in the SMF that arises due to the different number of low mass galaxies in each model alone.  For both the MC and H$_2$ models, we populate a SMF following the method outlined in \citet{GK2017} in which the scatter, $\sigma$, in the stellar mass at a given halo mass grows as  
\begin{equation}
\sigma = 0.2 + \gamma (\rm log_{10} M_{halo} - \rm log_{10} M_1)
\label{eq1}
\end{equation}
where $\gamma$ is the rate as which the scatter grows, and M$_1$ is a characteristic halo mass above which the scatter remains constant.  We adopt the scatter model for field galaxies from \citet{GK2017}, where M$_1 = 10^{11.5}$ M$_{\odot}$ and $\gamma = -0.25$.
We populate the $z=0$ ELVIS catalogs \citep{GarrisonKimmel2014} with this SMHM relation, and then assign galaxies as ``dark'' based on the fraction of luminous to non-luminous halos shown in Figure \ref{fig:dark} for each of our models.  The resulting SMFs are shown in Figure \ref{fig:mch2smf} as the blue (MC) and green (H$_2$) lines.  We do this $1000$ times in order to estimate our errors (dashed lines), and normalize the results so that each has 12 galaxies with stellar mass comparable to Fornax \citep[this is roughly the number of Fornax-mass galaxies within 1 Mpc][]{McConnachie2012}.  Figure \ref{fig:mch2smf} demonstrates that there is a difference in slope below M$_{star} \sim 10^5$ M$_{\odot}$, with the H$_2$ simulation predicting the shallower faint-end slope. 

The blue and green lines adopt the same slope and scatter of the SMHM relation, and thus only reflect the change in the fraction of ``occupied'' dark matter halos shown in Figure \ref{fig:dark}.  However, the H$_2$ model tends to form fewer stars in matched halos at the low mass end, which will yield a steeper SMHM slope that will impact the SMF.  As can be seen in Figure \ref{fig:hugeplot}, the \citet{GK2017} SMHM slope appears to be a decent fit to the central galaxies in the MC run (left panel).  However, the \citet{Brook2014} SMHM slope appears to be a better match to the central galaxies in the H$_2$ run.  Our simulations do not have enough galaxies to independently define the slope and scatter of the SMHM relation for each prescription, so we adopt the slope and scatter of \citet{GK2017} to describe the MC run, and \citet{Brook2014} to describe the H$_2$ run, in order to show the additional affect that the steeper SMHM relation will have on the SMF.  We assume the same growing scatter as in the \citet{GK2017} relation but applied to the \citet{Brook2014} slope, and again adopt the galaxy occupation fraction for the H$_2$ simulation.  The result is shown as the orange line in Figure \ref{fig:mch2smf}.  It can now be seen that there is a dramatic difference in the number of expected UFD galaxies in the two models.

The SMF for the two runs is indistinguishable above M$_{star} \sim 10^5$ M$_{\odot}$.  Since this is the mass range that is currently best constrained, current observations cannot constrain the two models.  However, galaxies at lower stellar masses are being discovered by surveys like DES and HSC-SSP, and potentially hundreds will be found near the Milky Way with the {\it Large Synoptic Survey Telescope} \citep[LSST;][]{tollerud08, Walsh2009,Newton2018}, and out to greater distances using integrated light surveys \citep{Danieli2018}.  The masses of the UFDs discovered in DES is highlighted by the grey region in Figure \ref{fig:mch2smf}. It can be seen that pure number counts of faint dwarfs in LSST (i.e., UFDs found out to $\sim$1 Mpc) can help us to constrain how star formation proceeded at high redshift.

\subsection{Satellites of Dwarf Galaxies}

\begin{figure*}
\centering
\includegraphics[width=1.5\columnwidth]{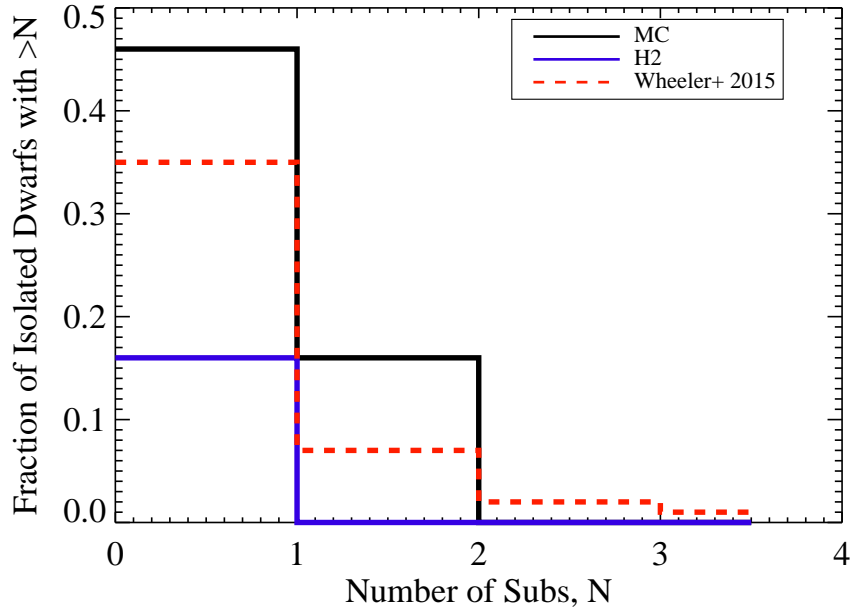}
\caption{Comparison of the predictions from hydrodynamic simulations for the number of luminous subhalos around dwarf galaxies.  The two solid lines are for the simulations in this work (black $=$ MC model, blue $=$ H$_2$ model), the red dashed for \citet{Wheeler2015}.  Each simulation has a different star formation and/or feedback recipe, with the model in \citet{Wheeler2015} being more similar to the MC run (discussed further in the text).  This figure demonstrates that the subgrid physics of the simulation affects the predictions for luminous subhalos, in some cases by a factor of $\sim$4. }
\label{fig:fire}
\end{figure*}

As discussed above, the MC simulation contains five times more satellites than the H$_2$ run at $z=0$. 
Here we use this result to estimate the predicted frequency of satellites around dwarfs in the Local Group in the two models.

We follow the methodology in \citet{Wheeler2015} and use the ELVIS suite of collisionless zoom-in simulations of Local Group-like environments \citep{GarrisonKimmel2014} combined with the results of our two simulations to predict the frequency of a satellite around a dwarf galaxy with M$_{vir} \sim 10^{10}$ M$_{\odot}$.  To summarize the procedure: 
(1) we select isolated dwarf galaxies from the ELVIS suite from both the isolated and paired Milky Way-mass halo simulations and (2) we estimate the frequency of subhalos with peak $M_{vir} \geq 10^{8}$ M$_{\odot}$ within 50 kpc of the dwarf host. 
Figure \ref{fig:fire} shows the results.

As expected, the MC model predicts that UFD satellites of dwarf galaxies are more abundant than the H$_2$ model.  The MC run produces non-negligible frequencies for finding at least one satellite around a dwarf host, 46\%. 
The H$_2$ model predictions are roughly a factor of 4 lower, at 16\%. 
The H$_2$ model also suppresses galaxy formation to the point that no Local Group dwarf with M$_{vir} \sim 10^{10}$ M$_{\odot}$ is expected to host more than one luminous satellite companion, while multiple UFD satellites are possible in the MC run.

The \citet{Wheeler2015} results adopt the FIRE 1 star formation and feedback prescription \citep{FIRE}.  FIRE 1 adopted a threshold for star formation of $>$ 100 $m_H$ cm$^{-3}$, while the updated FIRE 2 adopts a higher density threshold of $>$ 1000 $m_H$ cm$^{-3}$ \citep{FIRE2}.  
Both FIRE 1 and FIRE 2 use the equilibrium model from \citet{KrumholzGnedin2011} to determine the H$_2$ fraction in a given particle, which is used to calculate the self-shielding of the gas particle and the cooling rate.
The presence of H$_2$ for star formation is an explicit requirement, which is usually ensured by the simultaneous requirement of a high density threshold. However, because FIRE adopts the equilibrium model of H$_2$, there is no delay in star formation due to the H$_2$ formation time. 

Figure \ref{fig:fire} demonstrates that the predictions in \citet{Wheeler2015} are closer to our MC results than H$_2$ results.  This is in line with our expectations given the similarity in star formation threshold (and resolution) between the FIRE 1 model and the MC model.  Of course, both the feedback and reionization models in FIRE are different than those used here, which may also play some role in the results.  We note, though, that we have adopted a reionization model \citep{HM2012} that is known to lead to earlier reionization than that used in FIRE \citep{Faucher2009}, as was shown in \citet{Onorbe2017}.  Despite the stronger, earlier heating in our MC model, we predict slightly more satellites than \citet{Wheeler2015}.  

The results presented in Figure \ref{fig:fire} serve to underscore that current predictions for the number of UFD satellites expected around Local Group dwarfs should be approached with caution.  This work highlights the range of values we expect to see in state-of-the-art zoom simulations that reach the UFD mass/luminosity range.\footnote{ Though we note that these results are all currently based off of simulations of classical field dwarfs, and no one has yet simulated UFDs around a Milky Way-mass galaxy in a cosmological context.}  However, we do not currently know what star formation physics model is ``correct.''  Until future observations better constrain the models, we must find independent constraints from existing observations.

\section{Discussion}\label{discuss}

In this work, we have found that there is a significant delay in star formation and reduction in overall efficiency of star formation in simulated UFD galaxies when adopting a non-equilibrium H$_2$-based star formation prescription relative to a prescription that adopts a commonly-used temperature-density threshold.   We verified that the LW flux external to the galaxies in our H$_2$ model (due to either the UV background or nearby star forming galaxies) is not high enough to dissociate H$_2$ in the halos that fail to form stars.  Rather, the reduction in star formation in the H$_2$ model is due the long formation times of H$_2$ at low metallicities in low mass halos.  The delay in H$_2$ formation prevents or suppresses star formation in UFD galaxies because reionization can heat their gas before significant star formation can occur.  In contrast, the MC model allows star formation to occur as soon as gas reaches a high enough density threshold, and more stars form before reionization can remove the gas.

\subsection{Limitations of the Simulations}

It is unlikely that either of the explored models, MC or H$_2$, is ``correct.''  The MC model does not take into account whether the gas can shield when forming stars, and shielding of the gas is likely to be a necessary ingredient for star formation to occur.  Similarly, it has been argued that H$_2$ is not necessary for star formation, but that its presence is correlated with the ability of the gas to shield \citep{Glover2012,MacLow2012,Krumholz2011,Krumholz2012,Clark2014}. Hence, it is possible that neither model accurately reflects the physics of star formation in the first halos.  Instead, a model that links star formation to shielding may be more appropriate \citep{Byrne2019}. 

We emphasize that not all H$_2$ models for star formation will behave in the same way as our adopted non-equilibrium H$_2$ model.  As discussed throughout this paper, some H$_2$ models assume equilibrium between the formation and destruction of H$_2$ in the interstellar medium.  Because H$_2$ formation is not explicitly followed, there will be no dependence on the formation time of H$_2$ in equilibrium models.  Thus, equilibrium H$_2$ models may behave more like the MC model, though this remains to be tested.

\subsubsection{Resolution}

 As seen in Equation \ref{eq2}, the time scale for H$_2$ formation is density dependent.  Resolution will limit the maximum densities that we are capable of reaching.  Figure \ref{fig:phase} shows that our dwarf galaxies in the H$_2$ run that are able to form stars are able to reach gas densities of $1000 - 10^5 \,m_H$ cm$^{-3}$.  Because lower mass halos will contain fewer gas particles (i.e., fewer resolution elements), this limits their ability to reach the same high densities. Thus, it is possible that some stars should be forming in our dark halos, but that we are unable to capture the process.  

However, resolution is a limitation of all simulations, and our goal here is not to present the H$_2$ model as correct.  If we are missing star formation in lower mass halos, then we would recommend that simulators adopt a model that does not suppress star formation given their resolution (e.g., an equilibrium H$_2$ model, or a temperature/density threshold model).  

 It is not clear, though, that we are missing star formation in lower mass halos.  Our H$_2$ model suppresses star formation in halos below 10$^{8.5}$ M$_{\odot}$, and there is no indication to date that this is contrary to observations.   \citet{Jethwa2018} used abundance matching to put a lower limit on the peak halo masses of Milky Way UFDs of $> 2.4 \times$ 10$^8$ M$_{\odot}$. 
Meanwhile, \citet{Tollerud2018} suggest that reionization prevents galaxy formation in halos below 3$\times$10$^8$ M$_{\odot}$ in peak halo mass.  Their model is simple, with a sharp transition that has all halos hosting galaxies above this mass and none below it.  This behavior is actually quite similar to our H$_2$ model.  Thus, both of our models are in reasonable agreement with current observational data.


\subsubsection {Reionization model}

Many of our low mass halos in the H$_2$ model are not able to form stars before reionization prevents them from doing so. Hence, our results may be sensitive to reionization model. Reionization is expected to leave a visible imprint on the satellite luminosity function in the UFD range \citep{Bose2018}. We have adopted the same model in both simulations, following \citet{HM2012}.  However, \citet{HM2012} has been shown to heat the IGM earlier ($z \sim 15$) than it should \citep{Onorbe2017}, making the impact of reionization particularly strong on our results.  We have left it to future work  to quantify the impact of a gentler reionization model on the formation of ultra-faint dwarf galaxies, and restrict the focus of this paper to the role of star formation recipe alone.

 However, the main limitation of our reionization model is the fact that it is uniform throughout the simulation volume.  This is common for cosmological galaxy simulations, as the radiative transfer required to explicitly follow patchy reionization is too computationally expensive (and, as noted in Section \ref{Sims}, these runs already cost millions of CPU hours each).  Our dwarf volumes are $\sim$5 Mpc away from a Milky Way-mass galaxy, meaning that they may be in a lower density region that was not ionized as early as the higher density regions surrounding massive galaxies.  A more realistic reionization model for this region may then allow some of our dark halos to form stars.  The early reionization of \citet{HM2012} may be a better representation of what subhalos that fell into the Milky Way at early times experienced.  Because the goal of UFD modeling is often to make predictions for LSST, a model somewhere between the two extremes of late and early reionization is probably more appropriate.  However, an accurate study will require radiative transfer to follow the flux of LW radiation and the dissociation of H$_2$.


\subsection{Magellanic Cloud Satellites}

Recently, the {\it Dark Energy Survey} (DES), has nearly doubled the number of known UFDs in the Milky Way \citep{DES1,DES2, Kim2015a, Kim2015b, Koposov2015, Luque2017}. 
Many of the DES dwarfs are thought to potentially be satellites of the Magellanic Cloud system \citep{Deason2015, Yozin2015, Jethwa2016, Sales2017, Kallivayalil2018, Li2018}.
We note that we do not have an LMC-mass galaxy in this simulation volume.  The halo mass of the LMC is estimated to be at least M$_{halo} > 10^{11}$ M$_{\odot}$ \citep{Besla2015, Dooley2017b, Penarrubia2016, Laporte2018,Erkal2018}.  
A lot of work has been done recently to determine if the DES dwarfs are satellites of the LMC or Magellanic System \citep[e.g.,][]{Deason2015, Jethwa2016, Sales2017}.  The majority of studies find that 25-50\% of the newly discovered DES dwarfs are likely to have come in with the Magellanic Clouds. \citet{Kallivayalil2018} and \citet{Pace2018} derive proper motions from {\it Gaia} data to test whether the UFD satellites have kinematics consistent with falling in with the Clouds.  Between the two studies, seven UFDs are thought to be associated.  It is not immediately clear if these high numbers are compatible with our predictions, given our lack of LMC-mass galaxies.  

The models of \citet{Dooley2017b} predict $\sim$5-15 UFDs with M$_{star} > 3000$ M$_{\odot}$ (the lower limit for at least seven star particles in our well-matched galaxy sample) around an LMC-mass galaxy, and $\sim$2-9 around an SMC-mass galaxy.  These results are consistent with the number of UFDs that are potentially associated with the Magellanic Cloud system.  
However, the abundance matching results of \citet{Dooley2017a} put the SMC in a halo with M$_{halo} > 10^{11}$ M$_{\odot}$, more massive than our most massive simulated galaxy.  Based on their abundance matching results, a more direct comparison of our most massive simulated galaxies is to WLM, 
for which \citet{Dooley2017a} expect to find roughly one UFD companion.  This suggests that our results are generally consistent with the estimates in \citet{Dooley2017a,Dooley2017b}, as they should be as long as our simulation results are reasonably described by one of the SMHM relations that they explore.  However, it is unlikely that even our MC model would predict seven satellite companions around an LMC-mass galaxy. To reach such high numbers, the satellites must be associated with the whole Magellanic system (SMC+LMC), rather than just the LMC.

%

\section{Summary}\label{Summary}

In this work, we have used the highest resolution simulations to date of a cosmic volume of dwarf galaxies in order to examine the effect of star formation recipes on the formation of UFDs.  We run our volume with two different star formation recipes that are common in the literature: one with a temperature/density threshold for star formation, and another that requires the presence of H$_2$ for star formation.  The main differences that manifest between the two models occur in extremely low metallicity gas and are 1) the timescale over which star formation takes place and 2) the effective gas densities at which star particles form (the first above 100 $m_H$ cm$^{-3}$ and the H$_2$ primarily above 1000 $m_H$ cm$^{-3}$ due to gravitational collapse during H$_2$ formation and reduced dust shielding). 
We find that these differences lead to drastically different results for galaxy formation in halos with M$_{vir} < 10^9$ M$_{\odot}$.  

Broadly, the ability of the stars in the MC model to form earlier leads to more galaxies that can form in low mass halos.
When the H$_2$ dependency is introduced into the star formation model, however, gas in low-metallicity, low-mass halos is less likely to reach significant molecular fractions prior to reionization and gas may never reach densities high enough for dust-shielding to allow for substantial H$_2$.
As a result, many fewer low-mass halos in the H$_2$ run produce stars.  Even when they do produce stars, they form substantially less than their matched counterparts in the MC run, leading to a steeper SMHM relation and shallower faint-end SMF.

For the two models that we examine here, 
we find twice as many resolved galaxies form in the lower threshold (MC) run.  However, most of these are satellites.  
We also find that more satellites {\it survive} in the MC run, and we conjecture that feedback in the satellites somehow contributes to the delayed mergers/disruption of these satellites.
The combined effect leads to five times as many satellites in the MC run than in the H$_2$ run.  

We have convolved our results with the halos in the ELVIS simulation suite \citep{GarrisonKimmel2014} in order to make predictions for the number of UFD satellites around Local Group dwarf galaxies.  We find that the MC model predicts four times as many dwarfs should host at least one luminous satellite compared to the H$_2$ model, while the H$_2$ model predicts that no Local Group dwarfs should host more than one UFD satellite. 
Our MC model produces a similar prediction to that of \citet{Wheeler2015}, where the density threshold for star formation was also 100 $m_H$ cm$^{-3}$.


Our goal in this paper is not to figure out which model is correct.  Rather, our goal is to stress the need for caution in interpreting simulated UFD results.  As simulations push to ever-higher resolution and UFDs begin to be modeled for the first time in fully cosmological simulations run to $z=0$, the predictions for UFDs are likely to vary from model to model due to the assumptions adopted by varying modelers.  This is in contrast to model predictions in the classical dwarf galaxy mass range because classical dwarf galaxies are capable of self-regulating their star formation, reducing dependency on the adopted prescriptions in their resulting stellar masses.  As we push into the UFD mass range, we are simulating galaxies for the first time that can no longer self-regulate. 

However, we have highlighted a few future observables that will help to pinpoint the conditions for star formation in the first low mass halos.  Specifically, we have shown that the SMF and the number of UFD satellites around classical dwarf galaxies will be strongly impacted by star formation in UFDs during the reionization epoch.  The predicted SMF that can be probed by LSST is shallower in the UFD range for the model that restricts star formation to gas that has H$_2$.  This suggests that pure number counts of UFDs that we discover in the future can help to pinpoint the conditions required for the first star formation in  low mass halos. A better constraint on the slope and scatter of the SMHM relation at low masses will also help to constrain the models.


\acknowledgements

The authors would like to thank Sarah Loebman, Jillian Bellovary, and Dan Weisz for useful discussions relating to this paper.  FDM acknowledges funding from the VIDA fellowship.  AMB and FDM acknowledge support from HST-AR-13925. AMB acknowledges support from NSF-AST-1813871. Resources supporting
this work were provided by the NASA High-End Computing (HEC) Program through the NASA Advanced Supercomputing (NAS) Division at Ames Research Center.  EA acknowledges support from the National Science Foundation (NSF) Blue Waters Graduate Fellowship. This research is part of the Blue Waters sustained-petascale computing project, which is supported by the National Science Foundation (awards OCI-0725070 and ACI-1238993) and the state of Illinois. Blue Waters is a joint effort of the University of Illinois at Urbana-Champaign and its National Center for Supercomputing Applications.  This work was supported by a PRAC allocation NSF award number OCI-1144357.

\bibliography{bibref}
\clearpage

\end{document}